# Remarks on "Noncontextual Hidden Variables and Physical Measurements"

C.F. Boyle [*]

*University of Central England, Birmingham B42 2SU, U.K.*

and

R.L. Schafir [†]

*17 Milton Lodge, London N21 3NQ, U.K.*

**Abstract**:

Kent [1] has constructed a hidden variable theory by taking the finite precision of physical measurements into account. But its claim to noncontextuality has been queried [3-7], and it shown here that there is a particularly simple way of seeing why it cannot be noncontextual.

Kent [1] has constructed a hidden-variable theory by constructing a set of projective decompositions of the identity which is dense in the set of all projective decompositions of the identity, and in which just one projector in each decomposition has the value "true" and all others have the value "false". This is to show that it is possible to have a theory in which observables have pre-existing values if the finite precision of real measurement is taken into account; and in a follow-up article Clifton and Kent [2] set out to show that the relative frequencies of such a theory can coincide with the probabilities predicted by quantum mechanics.

A number of responses have now appeared to this paper [3-7], and in particular the claim that the theory is noncontextual has been queried. No argument is presented in Ref. [1] in support of this claim, though in Ref [2] there is a mention of the fact that the "true" projector can remain unchanged while changing the other projectors which are included in the decomposition alongside it. But if this is the idea of why the theory is noncontextual then it is insufficient, as we shall see.

---

[*] e-mail: conall.boyle@uce.ac.uk
[†] e-mail: shfeer@hotmail.com

Indeed, for the following reason alone, the theory cannot be noncontextual. Nonlocality, as Kent himself says, must still hold for his construction, since it has been verified by observational facts. Yet nonlocality is simply a particular kind of contextuality, when dealing with observables associated with spacelike-separated observers; it is inconsistent to claim that a theory is both nonlocal and noncontextual. A situation in which nonlocality occurs could be reinterpreted as a situation in which far-apart particles are replaced by different attributes of the same particle [3], and we would then say that contextuality has been proved, and could be demonstrated by suitable observations at the macroscopic level (so cannot be invalidated by finite precision).

The purpose of this comment is to point out that it can be seen very simply why the theory is not noncontextual, if we note a certain basic feature of the theory.

The feature is this. There is always a "true" projector close to *any* given projector (though there is also a "false" projector close to it as well). In quantum mechanics there are certain constraints on the possible values which can be observed for a set of commuting observables; but this over-permissiveness of Kent's construction allows *any* value to occur, since we can always find a "true" projector close to any other projector.

Now let us look at the conditions which are required for noncontextuality. First let us look at the exact case, i.e. when finite precision is not involved, and consider the simplest case of when contextuality versus noncontextuality can occur. Take three observables A, B and C such that $[A,B] = 0$ and $[A,C] = 0$, but $[B,C] \neq 0$ (which makes A necessarily degenerate), and take the two commuting sets as complete. For noncontextuality val(A) must be the same whether B or C is measured at the same time. So, writing $\alpha$, $\beta$ and $\gamma$ for val(A), val(B) and val(C), we have that the projector $P_\alpha$ into the $\alpha$-eigenspace of A is "true" and also that the eigenvectors of B and C corresponding to $\beta$ and $\gamma$ should lie inside the $\alpha$-eigenspace, which is also the eigenspace which is selected if A is measured on its own. So in terms of projectors which satisfy the conditions for being in Kent's constructed sets, for one decom-position of the identity consisting only of $P_\alpha$, $I - P_\alpha$, and two finer decompositions in which one includes $P_\beta$ and the other includes $P_\gamma$:

*If $P_\alpha$, $P_\beta$, $P_\gamma$ are true (in their respective decompositions), then $P_\alpha =$ the projector onto the subspace spanned by $|\beta\rangle$ and $|\gamma\rangle$.*

Note that more is involved than merely the requirement that the projector associated with the α-subspace has the same value whichever other states *outside* the subspace are considered.

Can this condition be incorporated into the exact theory? We know that it cannot, since the exact theory can be proved to be contextual. There must be some cases where the α-eigenspace which is obtained if B is measured at the same time, cannot occur if C is measured at the same time.

What answer can be given in the "finite precision" theory when we have such a case? [1] Presumably the answer is that we are no longer measuring exactly the observable intended, so when it comes to the alternative case (co-measuring C instead of B), we may be taking a slightly different degeneracy eigenspace with a slightly different state when the co-measured observable lifts the degeneracy. According to Kent's construction there are indeed "true" hypersurfaces close to the first hypersurface. But here the overpermissiveness of the theory comes into effect: there are "false" ones as well, and in general we can only take one of the true ones at the expense of getting a set of observed values which are not in accordance with quantum mechanics.

Perhaps the best way of regarding Kent's construction is as a confirmation the results of quantum mechanics can indeed be mimicked by hidden variable theories, at least provided that finite precision is taken into account. There already exists the Bohm theory of course, but in case any contradiction is found in that, or prediction at odds with quantum mechanics, here is an independent demonstration of the fact. But any such theory must be contextual and nonlocal, just as quantum mechanics itself is. And so must be any other theory which predicts the same observational results as quantum mechanics.

---

[1] The proofs of noncontextuality (see e.g. Mermin [8]) do not explicitly give a particular case, but give a choice that it must be one of several possibilities. But we can asssume for this argument that we have a case where it is known that a particular combination of values cannot occur.